# Improving the energy density and flexibility of PMN-0.3PT based piezoelectric generator by composite designing

Abhishek Kumar[1], Kaushik Das[2], Amritendu Roy[1, *]

[1] School of Minerals, Metallurgical and Materials Engineering, Indian Institute of Technology Bhubaneswar, Odisha-752050, India

[2] Department of Metallurgy and Materials Engineering, Indian Institute of Engineering Science and Technology Shibpur, Howrah-711103, West Bengal, India

[*]Corresponding author: amritendu@iitbbs.ac.in

## Abstract

Ceramics based piezoelectric generators are known for their high energy density and poor flexibility. In this work, $v_r$-PMN-0.3PT/PDMS 2-2 composite with optimum PMN-0.3PT content ($v_r$) was designed that demonstrated enhanced output energy density and superior mechanical flexibility under dynamic mechanical excitation. $v_r$-PMN-0.3PT/PDMS 2-2 composite with different PMN-PT reinforcement content ($v_r$) and two different reinforcement configurations were fabricated and characterized for effective electro-elastic properties and energy harvesting response. Parallelly, using the finite element method and analytical models, effective electromechanical properties were calculated. Composites with parallel connectivity of the reinforcement phase demonstrated enhanced piezoelectric charge coefficient ($d_{33}^c$) even with low PMN-0.3PT content whereas the relative permittivity ($\kappa_{33}^c$) and elastic modulus ($E_{33}^c$) exhibited a linearly increasing trend with reinforcement volume fraction. At a compressive load of 50 N and 5 Hz frequency, a piezoelectric generator (PG) based on a $v_r = $ 0.2, $v_r$- PMN-0.3PT/PDMS 2-2 composite with parallel connectivity produced a maximum short-circuit current density of 69 nA/cm$^2$ and an open-circuit electric field of 189 V/cm, translating to a maximum output power density of ~13 µW/cm$^3$ higher than that of pristine PMN-0.3 PT based piezoelectric generator. Estimated mechanical flexibility was found to be ~53 % higher than that of pristine PMN-0.3PT.

1. **Introduction**

Advanced wireless technologies and microelectronics have empowered a new generation of wearable technology, MEMS-based devices, sensors, actuators, etc. [1]. Furthermore, "Internet of Things" (IoT) has enabled the deployment of smart sensors in remote locations where it is challenging and occasionally impossible to charge batteries [2]. In this regard, high energy-density piezoelectric generators (PGs) appear promising to power such devices [3]. For these applications, to endure prolonged cyclic mechanical loading while in service, superior piezoelectric properties and mechanical flexibility are essential. Piezoelectric "figure-of-merit" analysis predicts that materials with high piezoelectric charge coefficient and low elastic compliance and dielectric permittivity are ideal for energy harvesting in compression mode of operation. Due to its large $d_{33}$ over other piezoceramics, PZT, and PMN-0.3PT are classic examples of efficient systems for energy harvesting [4,5].

Traditional piezoceramics, however, are naturally brittle and unreliable during cyclic mechanical loading [6]. The above attributes have led to the design of ceramic-polymer composites with impressive mechanical flexibility. Reasonable electro-elastic properties vis-à-vis energy output were reported with PZT/PVDF [7,8], PZT/PDMS [9–11], PMN-PT/PVDF [12,13], and PMN-PT/PDMS [14] 0-3 composites where ceramic particulates are randomly dispersed in the polymer matrix. However, the issue here is that at higher ceramic concentrations, the porosity content in the composite is large, leading to the chances of premature mechanical failure while in service. Furthermore, the output energy density is at least one order of magnitude smaller than the pristine ceramic, even with large ceramic content. This leaves ample opportunities to come up with novel composite designs with superior energy density with reasonable mechanical flexibility. In this regard, our earlier work suggested that the oriented configuration of the reinforcement phase in the composite could result in superior electro-elastic properties in piezoelectric-polymer composite [14].

Reinforcement orientation in the matrix could be achieved by specific designing strategies, such as with 1-3 and 2-2 configurations of the reinforcement and the matrix, respectively [15]. Indeed, composites with 1-3 and 2-2 configurations yield excellent piezoelectric performance [16–18]. Due to its easy fabrication steps and control over shape parameters, 2-2 piezoelectric composites are preferred over 1-3 composites. Fig. S1 in the supplementary information schematically portrays the 2-2 composite designing schemes. The majority of the previous research on 2-2 piezoelectric composites involves the estimation of relative permittivity ($\kappa$), piezoelectric charge coefficient ($d_{ij}$), electromechanical properties ($k_t, Q_m$), and damping performance (storage modulus, loss modulus, and loss factor) [19–28]. 2-2 PZT/epoxy piezoelectric composite [19] reported an increase in the piezoelectric charge coefficient $d_{33}$, relative permittivity $\kappa_{ii}^T$, thickness-electromechanical coupling coefficient ($k_t$) with the increase in the volume fraction of PZT reinforcement. Similar results were reported in 2-2 piezoelectric composites (PZT-5 as the piezoelectric active phase and cement-epoxy polymer as the passive phase) [29]. Earlier reports on 2-2 piezoelectric generator (device area = 480 mm$^2$) made from PIMNT (single crystal)/epoxy composites reported an open-circuit voltage of ~54 V and short-circuit current of 6.7 µA under bending mode [30]. PZN-PZT/PDMS and PZN-PZT/MW-CNTs/PDMS PEHs (1.5 cm × 1.5 cm × 0.1 cm) by Hao et al. [31] exhibited an output short-circuit current of ~200 nA and ~500 nA, respectively, demonstrating a significant potential for energy harvesting applications. In this regard, PMN-PT, with its class-leading electromechanical performance, could be an interesting proposition for developing high-energy density piezoelectric generators [5,14]. However, to the best of our knowledge, PMN-PT based 2-2 piezoelectric composite vis-à-vis effective elastic, dielectric, and piezoelectric properties as well as energy harvesting have not yet been investigated. In addition, no theoretical study has been conducted to assess the energy harvesting performance of the 2-2 piezoelectric composites.

Considering the above research gap, in the present work, we conduct a combined experimental and theoretical analysis to demonstrate the effect of PMN-0.3PT content on the electromechanical properties and energy harvesting performance of 2-2 PMN-0.3PT/PDMS piezoelectric composite. We show that the 2-2 parallel connectivity based $v_r$-PMN-0.3PT/PDMS-based composites ($v_r$ = 0.1, 0.20, and 0.43) demonstrate superior effective elastic, dielectric, and piezoelectric properties compared to the series connectivity through experiments as well as finite element-based models. The optimum composite composition appears to be $v_r$ = 0.20 in terms of effective properties. The energy harvesting performance of $v_r = 0.20$-based 2-2 parallel connected composite was demonstrated to be even higher than that of pristine PMN-0.3PT under identical loading condition albeit with superior mechanical flexibility.

## 2. Experimental and Computational details

### 2.1 Experiment

#### 2.1.1 Composite fabrication

The 2-2 piezoelectric composites can be represented in two different geometrical configurations: series and parallel connections. The process flowchart for the fabrication of the 2-2 PMN-0.3PT/PDMS composites with both series and parallel connectivities is presented in Fig. 1. Conventional solid-state reaction technique was used for synthesis of $0.7Pb(Mg_{1/3}Nb_{2/3}O_3) - 0.3PbTiO_3$ (PMN-0.3PT) ceramic [5]. For composite with parallel connectivity, ceramic-strips were cut from a sintered pellet. The dimensions of each strip are 12 mm × 2 mm × 1.5 mm. The volume fraction ($v_r$ = 0.10, 0.20, and 0.43) of PMN-0.3PT in the 2-2 composites was controlled by the number of PMN-0.3PT strips in the matrix. PMN-0.3PT ceramic-strip was put in the mold and PDMS (PDMS and the curing agent: 10:1) was added in the casting mold. The as-fabricated composite was cured at 80°C for 2 h. Subsequently, the composite was taken out of the mold and then cut into 15 mm × 15 mm pieces.

In the case of composite with series connectivity, PVDF was added as a binder to fabricate the PMN-PT layer. The cast film was subjected to drying at a temperature of 60°C for a duration of 4 h. Following this, a PDMS polymer mixture consisting of PDMS and a curing agent, in a ratio of 10:1 was applied onto the dried PMN-0.3PT film. Subsequently, a PMN-0.3PT layer was cast onto the PDMS film, which was then overlaid with an additional layer of PDMS. The volume fraction of PMN-0.3PT-PVDF in the 2-2 configured composites was managed by film thickness. $v_r$-PMN-0.3PT-PVDF/PDMS with $v_r$ = 0.25, 0.50, and 0.75 were successfully fabricated.

### 2.1.2 Characterization technique

For electrical characterization, a layer of conductive silver paint (Sigma-Aldrich) was uniformly coated on both surfaces of the composite film (12 mm × 12 mm). The capacitance of the composite sample was measured at a frequency of 1 kHz and with a 1 V AC signal using the Wayne Kerr 6500B high-precision impedance analyzer. The relative permittivity (($\kappa_{33}^c$) was calculated from the capacitance: $\kappa_{33}^c = Ct/\kappa_0 A$, where A represents the surface area of the electrode, $t$ denotes the thickness of the composite sample, and $\kappa_0$ corresponds to the permittivity of the vacuum. Impedance measurements were conducted on poled composite samples (1 kV/mm) over the frequency range of 20 Hz to 1 MHz at room temperature. The thickness coupling coefficient ($k_t^c$) of the piezoelectric composites can be mathematically represented in terms of effective material constants as [32]:

$$k_t^c = \frac{d_{33}^c}{\sqrt{S_{33}^c \times \varepsilon_{33}^c}} \qquad (1)$$

or, the value of $k_t^c$ can be determined by measuring the resonance frequency $f_{r\,(t)}$ and anti-resonance frequency ($f_{a\,(t)}$) of a piezoelectric composite using the relation [33]:

$$k_t^c = \sqrt{\frac{\pi}{2} \times \frac{f_{r\,(t)}}{f_{a\,(t)}} \times \tan\left(\frac{\pi}{2} \times \frac{f_{a\,(t)} - f_{r\,(t)}}{f_{a\,(t)}}\right)} \qquad (2)$$

Eq. 1 and 2 were used for calculation of the elastic modulus ($E_{33}^c = C_{33}^c = \frac{1}{S_{33}^c}$) of the 2-2 $v_r$-PMN-0.3PT/PDMS ($v_r$ = 0.10, 0.20, and 0.43) piezoelectric composites. Piezoelectric charge coefficient ($d_{33}$) of the poled samples was measured using $d_{33}$ meter (Model: D33PZO1) at 110 Hz frequency. Copper wires were affixed on both sides of the sample, and subsequently laminated to mitigate any potential triboelectric contribution to the output energy. Using a high resistance Keithley 6517B electrometer, the output voltage and current response of piezoelectric generator (made from the composite) were measured under various cyclic compressive loads at different frequencies.

## 2.2 Computation

### 2.2.1 Analytical Model

The piezoelectric constitutive relationships that exhibit coupling characteristics between mechanical properties (stress and strain) and electrical properties (electric displacement and electric field) are given in matrix form as, [34]

$$\begin{bmatrix} \sigma \\ D \end{bmatrix} = \begin{bmatrix} C^E & -e^T \\ e & \kappa^\varepsilon \end{bmatrix} \begin{bmatrix} \varepsilon \\ E \end{bmatrix} \qquad (3)$$

where, $\sigma, \varepsilon, E, D, C^E, \kappa^\varepsilon$, and $e$ are the stress tensor, strain tensor, electric field vector, electric displacement vector, elastic stiffness tensor at constant electric field, permittivity tensor at constant or zero strain, and piezoelectric coupling tensor, respectively. The superscript $T$ stands for the transpose of the matrix. The above constitutive equations represent several independent material properties for the PMN-0.3PT and PDMS material, such as elastic modulus, piezoelectric charge coefficient, and relative permittivity. Hence, all these material properties need to be known for the complete characterization of a 2-2 PMN-0.3PT/PDMS piezoelectric composite.

The 2–2 PMN-0.3PT/PDMS piezoelectric composite with parallel connectivity (illustrated in Fig. 2(a)) possesses an interface between the PDMS and PMN-0.3PT normal to the 1-direction. The relationships between the stresses, strains, electric displacements, and the electric fields among PDMS matrix, PMN-0.3PT reinforcement, and PMN-0.3PT/PDMS composite are summarized in Table 1.

**Table 1.** The relationships between the stress and strain as well as electric displacement and the electric field in the PMN-0.3PT ($r$), PDMS ($m$), and PMN-0.3PT/PDMS composites ($c$) for the 2-2 layered piezoelectric composite. Voigt notation was used in conversion of two indices to a single index wherever applicable.

| Stresses | Strains |
|---|---|
| $\sigma_1^c = \sigma_1^m = \sigma_1^r$ | $\varepsilon_1^c = v_m \varepsilon_1^m + v_r \varepsilon_1^r$ |
| $\sigma_2^c = v_m \sigma_2^m + v_r \sigma_2^r$ | $\varepsilon_2^c = \varepsilon_2^m = \varepsilon_2^r$ |
| $\sigma_3^c = v_m \sigma_3^m + v_r \sigma_3^r$ | $\varepsilon_3^c = \varepsilon_3^m = \varepsilon_3^r$ |
| $\sigma_4^c = v_m \sigma_4^m + v_r \sigma_4^r$ | $\varepsilon_4^c = \varepsilon_4^m = \varepsilon_4^r$ |

$$\sigma_5^c = \sigma_5^m = \sigma_5^r \qquad\qquad \varepsilon_5^c = v_m\varepsilon_5^m + v_r\varepsilon_5^r$$

$$\sigma_6^c = \sigma_6^m = \sigma_6^r \qquad\qquad \varepsilon_6^c = v_m\varepsilon_6^m + v_r\varepsilon_6^r$$

| Electric displacement | Electric fields |
|---|---|
| $D_1^c = D_1^m = D_1^r$ | $E_1^c = v_m E_1^m + v_r E_1^r$ |
| $D_2^c = v_m D_2^m + v_r D_2^r$ | $E_2^c = E_2^m = E_2^r$ |
| $D_3^c = v_m D_3^m + v_r D_3^r$ | $E_3^c = E_3^m = E_3^r$ |

Analytical model by Kar-Gupta et al. [35] to predict the effective properties of periodic 2-2 multi-layered composites captures the complete electromechanical response of 2-2 piezoelectric composite. Using the constitutive equations as well as the relationships mentioned in Tables 1, all the independent material constants (81) of the 2-2 piezoelectric composite with parallel connectivity can be calculated by using:

$$\begin{bmatrix} C_{11}^c & C_{12}^c & C_{13}^c & C_{14}^c & C_{15}^c & C_{16}^c & -e_{11}^c & -e_{21}^c & -e_{31}^c \\ C_{21}^c & C_{22}^c & C_{23}^c & C_{24}^c & C_{25}^c & C_{26}^c & -e_{12}^c & -e_{22}^c & -e_{32}^c \\ C_{31}^c & C_{32}^c & C_{33}^c & C_{34}^c & C_{35}^c & C_{36}^c & -e_{13}^c & -e_{23}^c & -e_{33}^c \\ C_{41}^c & C_{42}^c & C_{43}^c & C_{44}^c & C_{45}^c & C_{46}^c & -e_{14}^c & -e_{24}^c & -e_{34}^c \\ C_{51}^c & C_{52}^c & C_{53}^c & C_{54}^c & C_{55}^c & C_{56}^c & -e_{15}^c & -e_{25}^c & -e_{35}^c \\ C_{61}^c & C_{62}^c & C_{63}^c & C_{64}^c & C_{65}^c & C_{66}^c & -e_{16}^c & -e_{26}^c & -e_{36}^c \\ e_{11}^c & e_{12}^c & e_{13}^c & e_{14}^c & e_{15}^c & e_{16}^c & \kappa_{11}^c & \kappa_{12}^c & \kappa_{13}^c \\ e_{21}^c & e_{22}^c & e_{23}^c & e_{24}^c & e_{25}^c & e_{26}^c & \kappa_{21}^c & \kappa_{22}^c & \kappa_{23}^c \\ e_{31}^c & e_{32}^c & e_{33}^c & e_{34}^c & e_{35}^c & e_{36}^c & \kappa_{31}^c & \kappa_{32}^c & \kappa_{33}^c \end{bmatrix} = [R]^T[A] + [B] \qquad (4)$$

Where [R], [A] and [B] are expressed as:

$$[R] = [P]^{-1}[Q] \qquad (5)$$

$$[P] = \begin{bmatrix} v_r C_{11}^m + v_m C_{11}^r & v_r C_{15}^m + v_m C_{15}^r & v_r C_{16}^m + v_m C_{16}^r & -(v_r e_{11}^m + v_m e_{11}^r) \\ v_r C_{15}^m + v_m C_{15}^r & v_r C_{55}^m + v_m C_{55}^r & v_r C_{56}^m + v_m C_{56}^r & -(v_r e_{15}^m + v_m e_{15}^r) \\ v_r C_{16}^m + v_m C_{16}^r & v_r C_{56}^m + v_m C_{56}^r & v_r C_{66}^m + v_m C_{66}^r & -(v_r e_{16}^m + v_m e_{16}^r) \\ v_r e_{11}^m + v_m e_{11}^r & v_r e_{15}^m + v_m e_{15}^r & v_r e_{16}^m + v_m e_{16}^r & v_r \kappa_{11}^m + v_m \kappa_{11}^r \end{bmatrix} \qquad (6)$$

$$[Q] = \begin{bmatrix} C_{11}^r & v_r(C_{12}^r - C_{12}^m) & v_r(C_{13}^r - C_{13}^m) & v_r(C_{14}^r - C_{14}^m) & C_{15}^r & C_{16}^r & -e_{11}^r & -v_r(e_{21}^r - e_{21}^m) & -v_r(e_{31}^r - e_{31}^m) \\ C_{15}^r & v_r(C_{25}^r - C_{25}^m) & v_r(C_{35}^r - C_{35}^m) & v_r(C_{45}^r - C_{45}^m) & C_{55}^r & C_{56}^r & -e_{15}^r & -v_r(e_{25}^r - e_{25}^m) & -v_r(e_{35}^r - e_{35}^m) \\ C_{16}^r & v_r(C_{26}^r - C_{26}^m) & v_r(C_{36}^r - C_{36}^m) & v_r(C_{46}^r - C_{46}^m) & C_{56}^r & C_{66}^r & -e_{16}^r & -v_r(e_{26}^r - e_{26}^m) & -v_r(e_{36}^r - e_{36}^m) \\ e_{11}^r & v_r(e_{12}^r - e_{12}^m) & v_r(e_{13}^r - e_{13}^m) & v_r(e_{14}^r - e_{14}^m) & e_{15}^r & e_{16}^r & k_{11}^r & v_r(k_{12}^r - k_{12}^m) & v_r(k_{13}^r - k_{13}^m) \end{bmatrix} \quad (7)$$

$[A] =$

$$\begin{bmatrix} C_{11}^m & v_m(C_{12}^m - C_{12}^r) & v_m(C_{13}^m - C_{13}^r) & v_m(C_{14}^m - C_{14}^r) & C_{15}^m & C_{16}^m & e_{11}^m & v_f(e_{21}^m - e_{21}^r) & v_r(e_{31}^m - e_{31}^r) \\ C_{15}^m & v_m(C_{25}^m - C_{25}^r) & v_m(C_{35}^m - C_{35}^r) & v_m(C_{45}^m - C_{45}^r) & C_{55}^m & C_{56}^m & e_{15}^m & v_f(e_{25}^m - e_{25}^r) & v_r(e_{35}^m - e_{35}^r) \\ C_{16}^m & v_m(C_{26}^m - C_{26}^r) & v_m(C_{36}^m - C_{36}^r) & v_m(C_{46}^m - C_{46}^r) & C_{56}^m & C_{66}^m & e_{16}^m & v_f(e_{26}^m - e_{26}^r) & v_r(e_{36}^m - e_{36}^r) \\ -e_{11}^m & -v_m(e_{12}^m - e_{12}^r) & -v_m(e_{13}^m - e_{13}^r) & -v_m(e_{14}^m - e_{14}^r) & -e_{15}^m & -e_{16}^m & k_{11}^m & v_f(k_{12}^m - k_{12}^r) & v_r(k_{13}^m - k_{13}^r) \end{bmatrix}$$

(8)

$[B]$

$$= \begin{bmatrix} 0 & C_{12}^r & C_{13}^r & C_{14}^r & 0 & 0 & 0 & e_{21}^r & e_{31}^r \\ C_{12}^m & v_m C_{22}^m + v_r C_{22}^r & v_m C_{23}^m + v_r C_{23}^r & v_m C_{24}^m + v_r C_{24}^r & C_{25}^m & C_{26}^m & e_{12}^m & v_m e_{22}^m + v_r e_{22}^r & v_m e_{32}^m + v_r e_{32}^r \\ C_{13}^m & v_m C_{23}^m + v_r C_{23}^r & v_m C_{33}^m + v_r C_{33}^r & v_m C_{34}^m + v_r C_{34}^r & C_{35}^m & C_{36}^m & e_{13}^m & v_m e_{23}^m + v_r e_{23}^r & v_m e_{33}^m + v_r e_{33}^r \\ C_{14}^m & v_m C_{24}^m + v_r C_{24}^r & v_m C_{34}^m + v_r C_{34}^r & v_m C_{44}^m + v_r C_{44}^r & C_{45}^m & C_{46}^m & e_{14}^m & v_m e_{24}^m + v_r e_{24}^r & v_m e_{34}^m + v_r e_{34}^r \\ 0 & C_{25}^r & C_{35}^r & C_{45}^r & 0 & 0 & 0 & e_{25}^r & e_{35}^r \\ 0 & C_{26}^r & C_{36}^r & C_{46}^r & 0 & 0 & 0 & e_{26}^r & e_{36}^r \\ 0 & -e_{12}^r & -e_{13}^r & -e_{14}^r & 0 & 0 & 0 & \kappa_{12}^r & \kappa_{13}^r \\ -e_{21}^m & -(v_m e_{22}^m + v_r e_{22}^r) & -(v_m e_{23}^m + v_r e_{23}^r) & -(v_m e_{24}^m + v_r e_{24}^r) & -e_{25}^m & -e_{26}^m & \kappa_{12}^m & v_m \kappa_{22}^m + v_r \kappa_{22}^r & v_m \kappa_{23}^m + v_r \kappa_{23}^r \\ -e_{31}^m & -(v_m e_{32}^m + v_r e_{32}^r) & -(v_m e_{33}^m + v_r e_{33}^r) & -(v_m e_{34}^m + v_r e_{34}^r) & -e_{35}^m & -e_{36}^m & \kappa_{13}^m & v_m \kappa_{23}^m + v_r \kappa_{23}^r & v_m \kappa_{33}^m + v_r \kappa_{33}^r \end{bmatrix}$$

(9)

For 2-2 composite with series connectivity, however, the above model needs to be modified as per the current experimental setup where direction 3 is the poling direction [35]. Let us suppose that the elastic, dielectric, and piezoelectric tensors are specified in a basis $\{u_1, u_2, u_3\}$ (Fig. 2(b)), and we want to calculate these tensors in another basis, $\{v_1, v_2, v_3\}$ (Fig. 2 (c)). The transformation tensor $(\Omega_{ij})$, written in matrix form as,

$$[\Omega_{ij}] = \begin{bmatrix} v_1 \cdot u_1 & v_1 \cdot u_2 & v_1 \cdot u_3 \\ v_2 \cdot u_1 & v_2 \cdot u_2 & v_2 \cdot u_3 \\ v_3 \cdot u_1 & v_3 \cdot u_2 & v_3 \cdot u_3 \end{bmatrix} \quad (10)$$

Note that, $v_1 \cdot u_3 = 1$ ($v_1 \parallel u_3$), $v_2 \cdot u_2 = 1$ ($v_2 \parallel u_2$), and $v_3 \cdot u_1 = 1$ ($v_3 \parallel u_1$). Therefore,

$[\Omega_{ij}] = \begin{bmatrix} 0 & 0 & 1 \\ 0 & 1 & 0 \\ 1 & 0 & 0 \end{bmatrix}$. The elastic, dielectric, and piezoelectric tensors are transformed as,

$$C_{ijkl}^{new} = \Omega_{ip}\Omega_{jq}\Omega_{kr}\Omega_{ls}C_{pqrs}^{old} \tag{11}$$

$$k_{ij}^{new} = \Omega_{ip}\Omega_{jq}k_{pq}^{old} \tag{12}$$

$$e_{ijk}^{new} = \Omega_{ip}\Omega_{jq}\Omega_{kr}e_{pqr}^{old} \tag{13}$$

The above formulation was used to transform the elastic, dielectric, and piezoelectric tensors into any new coordinate system. In the present work, firstly, the materials properties were transformed based on Kar-Gupta's assumptions (along 1-direction). The transformed electromechanical properties are presented in Table S3 of the supplementary section. The transformed electromechanical properties of PDMS and PMN-PT were used to estimate the effective properties of the composites with series connectivity, followed by further transformation along 3- direction.

### 2.2.2 Finite Element Model

As demonstrated in the following section, the 2-2 composite with parallel connectivity is ideal from energy harvesting perspective. Therefore, finite element analysis with kinematic uniform boundary conditions was employed only for the composites with parallel connectivity as implemented in COMSOL Multiphysics software [36]. Initially, a 2-2 parallel connectivity piezoelectric structure was created, based on the predetermined volume fraction and layer interface orientation. The kinematic or homogeneous boundary conditions and expressions used for determination of the effective relative permittivity, elastic modulus, and piezoelectric charge coefficient can be found in the supplementary information (Table S4). The boundary conditions (in the form of prescribed displacements and prescribed electric potentials) applied to the six surfaces of the piezoelectric device are also given in Table S4 [37]. The displacement components along the $x$, $y$, and $z$ coordinate axes are denoted as $u$, $v$, and $w$, respectively. $\varepsilon_0$ and $V_0$ represent the applied strain and applied electric potential, respectively. The components of volume-averaged stress ($\bar{\sigma}_{ij} = \frac{1}{V} \int \sigma_{ij}\, dV$), volume-averaged strain ($\bar{\varepsilon}_{ij} == \frac{1}{V} \int \varepsilon_{ij}\, dV$), volume-averaged electric field ($\bar{E}_i = \frac{1}{V} \int E_i\, dV$), and volume-averaged electric displacement ($\bar{D}_i = \frac{1}{V} \int D_i\, dV$) due to the application of strain and electric field were evaluated. Subsequently, the effective electromechanical properties of the composite were determined from the resultant volume-averaged stress, volume-averaged strain, volume-averaged electric field, and volume-averaged electric displacement derived from the FEM. A maximum reinforcement volume fraction of 0.53 was attained using the configuration shown in Fig. 3 (a)–(e). The volume fraction was varied by increasing or decreasing the number of reinforcement strips. The physics-controlled meshed composite with $v_r = 0.53$ is shown in Fig. 3 (f).

The output voltage and output current of the 2-2 PMN-0.3PT/PDMS piezoelectric composite (15 mm × 15 mm × 1.5 mm) with parallel connectivity were simulated using the COMSOL Multiphysics. The lower surface was fixed-constrained (solid mechanics) and grounded (electrostatics). The boundary load (5 N, 10 N, 20 N, 30 N, and 50 N) at a frequency of 1 Hz and 5 Hz was applied to the upper surface of the piezoelectric composite. Moreover, a floating potential boundary was applied to the upper surface. For meshing three-dimensional models, tetrahedral elements were utilized. A direct solver with a relative tolerance of 0.01 was employed. The simulations were conducted using a time-dependent approach for 2 second with a time step of 0.1 seconds.

## 3. Results and discussion

### 3.1 Effective elastic, dielectric, and electromechanical properties

Measured and computed effective elastic modulus ($E_{33}^c$), relative permittivity ($\kappa_{33}^c$), and piezoelectric charge coefficient ($d_{33}^c$) of 2-2 PMN-0.3PT/PDMS piezoelectric composites are plotted in Fig. 4 (a)–(c) as a function of reinforcement content and alignment of the reinforcement viz., series and parallel connectivity. Effective properties are normalized with respect to the reinforcement value. Although the effective elastic modulus for the composite with series connectivity was not possible experimentally. Fig. 4 (a) nonetheless, theoretical prediction using analytical models (discussed in section 2.2.1) shows that with parallel connectivity, one attains larger $E_{33}^c$ over series connectivity. The effective elastic modulus ($E_{33}^c$) of the 2-2 composites in parallel and series connectivities can also be predicted using models by Voigt [38] and Reuss [39], represented respectively, as $E_{33}^c = v_r E_{33}^r + v_m E_{33}^m$ and $E_{33}^c = \frac{E_{33}^r E_{33}^m}{v_r E_{33}^m + v_m E_{33}^r}$; where, $E_{33}^c$, $E_{33}^r$, and $E_{33}^m$ are the elastic modulus of PMN-0.3PT/PDMS composite, PMN-0.3PT reinforcement (7.6 × 10$^4$ MPa, calculated from impedance [5]), and PDMS matrix (1.92 MPa, calculated from tensile test [14]), respectively. $v_r$ and $v_m$ are the PMN-0.3PT ceramic and PDMS polymer volume fractions in the composite, respectively. Since PMN-0.3PT has a significantly higher elastic modulus than PDMS, the elastic modulus of 2-2 PMN-0.3PT/PDMS composites increases with increasing PMN-0.3PT content. There is, however, only a marginal modification in the elastic modulus of the composite with series connectivity with an increase in the PMN-0.3PT content. In contrast, a substantial increase is observed with parallel connectivity. The Reuss and modified Kar-Gupta models are in perfect agreement with each other when it comes to the series connectivity of 2-2 PMN-0.3PT/PDMS composites. For the parallel connectivity, the normalized effective elastic modulus modeled by both Kar-Gupta and FEM matches with the experimental results for all the PMN-PT volume fractions; however, the Voigt model overestimated the experimental results, as can be seen in Fig. 4 (a).

Relative permittivity ($\kappa_{33}^c$) of the 2-2 parallel connectivity ($v_r$ = 0.10, 0.20, and 0.43) and series connectivity ($v_r$ = 0.25, 0.50, and 0.75) piezoelectric composites were calculated using, $\kappa_{33}^c = \frac{Ct}{\kappa_0 A}$; where $C$ is the capacitance measured at 1 kHz, $A$ is the electrode area on the sample, $t$ is the sample thickness, and $\kappa_0$ is the free-space permittivity ($8.85 \times 10^{-12}$ F/m). Fig. 4 (b) plots the variation in the experimental and calculated normalized relative permittivity ($\kappa_{33}^c/\kappa_{33}^r$) of the 2-2 composite with series and parallel connectivity as a function of PMN-0.3PT content. In addition to the analytical model presented above, the effective relative permittivity ($\kappa_{33}^c$) of the two-phase composite can also be calculated by using the Wiener model [40] for parallel and series connectivities, respectively, using $\kappa_{33}^c = v_r \kappa_{33}^r + v_m \kappa_{33}^m$ and $\kappa_{33}^c = \frac{\kappa_{33}^r \kappa_{33}^m}{v_r \kappa_{33}^m + v_m \kappa_{33}^r}$ ; where, $\kappa_{33}^c$, $\kappa_{33}^r$, and $\kappa_{33}^m$ are the relative permittivity of PMN-0.3PT/PDMS composite, PMN-0.3PT reinforcement and PDMS matrix, respectively. $v_r$ and $v_m$ are the PMN-0.3PT ceramic and PDMS polymer volume fractions in the composite, respectively. PMN-0.3PT ceramic possesses a higher relative permittivity compared to the PDMS polymer matrix. Available theoretical models, including the one by Wiener [34], suggest that with an increase in the PMN-0.3PT content, there is a marginal increase in the relative permittivity ($\kappa_{33}^c$) with the series connectivity while a substantial increase in the $\kappa_{33}^c$ could be obtained for parallel connectivity. For the series connectivity, both Wiener and modified Kar-Gupta's normalized relative permittivity matches the experimental results for all the PMN-PT volume fractions. Both the Finite Element Method (FEM) and the Kar-Gupta analytical model for parallel connectivity yield identical predictions of $\kappa_{33}^c$ of PMN-0.3PT/PDMS composites across all the volume fractions of PMN-0.3PT. For the parallel connectivity, FEM and Kar-Gupta predicted $\kappa_{33}^c$ at $v_r = 0.20$ composite has a good agreement with the experimentally measured $\kappa_{33}^c$ as shown in Fig. 4 (b). However, for $v_r = 0.43$, all the models overestimate the experimental $\kappa_{33}^c$. low $\kappa_{33}^c$ may arise in the sample due to poor interfacial bonding between reinforcement. A similar observation was reported in 2-2 PZT/epoxy piezoelectric composites [19].

Fig. 4 (c) plots calculated and experimentally measured normalized piezoelectric charge coefficient, $d_{33}^c/d_{33}^r$ of 2-2 PMN-0.3PT/PDMS composite with series and parallel connectivity as a function of PMN-0.3PT volume fraction. All the composite samples are poled in direction-3. Fig. 4(c) demonstrates that the predicted $d_{33}^c$ of the composite with series connectivity does not show a noticeable increase even at a very high PMN-0.3PT volume fraction while in parallel connection, $d_{33}^c$ reaches close to the $d_{33}^r$(PMN-0.3PT) even with small PMN-PT content. To corroborate our theoretical observation, the effective piezoelectric charge coefficient of the composite as a function of reinforcement volume fraction was also estimated using the model by Newnham et al. [41]. In this model, the effective piezoelectric charge coefficient of the parallel composite depends on the elastic compliance and piezoelectric charge coefficient of the reinforcement and matrix phases; however, in series composites, it depends on the relative permittivity and piezoelectric charge coefficient of the reinforcement and matrix phases,

$$d_{33}^c = \frac{v_r d_{33}^r s_{33}^m + v_m d_{33}^m s_{33}^r}{v_r s_{33}^m + v_m s_{33}^r} \qquad (14)$$

$$d_{33}^c = \frac{v_r d_{33}^r \kappa_{33}^m + v_m d_{33}^m \kappa_{33}^r}{v_r \kappa_{33}^m + v_m \kappa_{33}^r} \qquad (15)$$

Where, $d_{33}^c$, $d_{33}^r$, and $d_{33}^m$ are the piezoelectric charge coefficient of PMN-0.3PT/PDMS composite, PMN-0.3PT reinforcement ($d_{33}^r$ = 237 pC/N), and PDMS matrix ($d_{33}^m$ = 0 pC/N), respectively. Also, $v_r$ and $v_m$ are the PMN-0.3PT ceramic and PDMS polymer volume fractions in the composite, respectively. For the series connectivity, both Newnham and modified Kar-Gupta's normalized piezoelectric charge coefficient matches the experimental results for all the PMN-PT volume fractions. For the parallel connectivity, the piezoelectric charge coefficient predicted by Newnham, Finite Element Method (FEM), and Kar-Gupta analytical model are identical for all PMN-0.3PT volume fractions (close to the $d_{33}^r$). Piezoelectric charge coefficient $d_{33}^c$ of the series and parallel connectivity 2-2 $v_r$-PMN-0.3PT/PDMS composites was measured using a $d_{33}$ piezometer. The Piezoelectric charge coefficient of the $v_r$ = 0.10, 0.20, and 0.43 parallel connectivity composite samples are found to be 183 pC/N, 194.5 pC/N, and 200.5 pC/N, respectively, which are approximately 81.5 %, 86.5 %, and 89.1 % of the $d_{33}^r$ values, respectively. However, for series connectivity, the composites with $v_r$ = 0.25, 0.50, and 0.75 demonstrate very low piezoelectric charge coefficients of 2.9 pC/N, 3.3 pC/N, and 3.1 pC/N, respectively. These data validate theoretical predictions [33][14]. Thus, in the subsequent study, we selected 2-2 PMN-0.3PT/PDMS with a parallel connection where the piezoelectric phase is poled in a 3-direction.

### 3.2 Mechanical flexibility

Clearly, 2-2 PMN-0.3PT/PDMS composite with parallel connectivity demonstrates large piezoelectric charge coefficient compared to the 2-2 composite with series connectivity as well as 0-3 composite [42]. However, it should also be noted that 0-3 PMN-0.3PT/PDMS composite demonstrates large mechanical flexibility which is critical for long service life [43]. In this regard, assessing the mechanical flexibility of 2-2 PMN-0.3PT/PDMS composite with parallel connectivity is important before putting it for device fabrication and characterization. The flexibility figure of merit ($f_{FOM}$) can be estimated by: $f_{FOM} = \frac{\sigma_y}{E}$, where $\sigma_y$ and E represent yield stress and elastic modulus, respectively [44]. The yield stress of the 2-2 PMN-0.3PT/PDMS composite ($\sigma_{33}^c$) with parallel connectivity was calculated by the rule of mixture [45], $\sigma_{33}^c = v_r \sigma_{33}^r + v_m \sigma_{33}^m$, where $\sigma_{33}^r$ and $\sigma_{33}^m$ represent the yield stress of PMN-0.3PT reinforcement (53.13 MPa [46]) and PDMS matrix (0.88 MPa [14]), respectively. The $f_{FOM}$ of 1.07 × 10$^{-3}$ was obtained for $v_r$ = 0.20 PMN-0.3PT/PDMS composites, representing an increase of ~53 % compared to the reported $f_{FOM}$ in PMN-PT [5]. This result confirms the enhanced flexibility of the 2-2 PMN-0.3PT/PDMS composite with parallel connectivity than the PMN-PT.

### 3.3 Energy harvesting performance

The output voltage ($V$) and output charge ($Q$) from a piezoelectric generator (PGs) under dynamic compressive load can be theoretically estimated using: $V = \frac{d_{33} \times F \times t}{\kappa_r \kappa_0 A}$ and $Q = d_{33} \times F$, where $d_{33}$ represents the piezoelectric charge coefficient, $F$ is the excitation force, $A$ is the device area, $t$ is the device thickness, and $\kappa_r$ denotes the dielectric permittivity of the material [47]. These relationships indicate that the output voltage and output charge are directly proportional to the magnitude of force excitation. The 2-2 $v_r$-PMN-0.3PT/PDMS ($v_r$ = 0.10, 0.20, and 0.43) parallel connectivity composites-based PGs were evaluated for their ability to harvest mechanical energy by applying a cyclic compressive load of 5-50 N at 1-5 Hz frequency. The short-circuit current density ($J_{sc}$) and open-circuit electric field ($E_{OC}$) under different compressive loads at frequency, 5 Hz are shown in Fig. 5 (a) and (b). It can be noted that $J_{sc}$ and $E_{OC}$ generated from $v_r$ = 0.20 under compressive load of 5 N at 5 Hz frequency are 7.44 nA/cm² and 28.5V/cm, respectively. These values increased to 68.75 nA/cm² and 189 V/cm when the load was increased to 50 N, confirming the dependence on the applied load [48–50]. Similar load-dependence behavior was observed in $v_r$ = 0.10 and $v_r$ = 0.43 composites, as well. Furthermore, the above correlations indicate an inverse relationship between voltage ($V$) and relative permittivity for constant $d_{33}$, while $Q$ is directly proportional to $d_{33}$. Relative permittivity monotonically increases with the PMN-PT volume fraction as shown in Fig. 4 (b). However, Fig. 4 (c) clearly shows that even at a very low PMN-0.3PT content, $d_{33}^c$ reaches close to the $d_{33}^r$ (PMN-0.3PT). Further increase in the PMN-PT content only cause marginal increase in the experimental $d_{33}^c$ value. Therefore, it can be inferred that beyond a specific PMN-PT content, there would be a drop in the output voltage. Fig. 7 shows that there is an initial increase in the experimental $J_{sc}$ and $E_{OC}$ with PMN-PT concentration. However, it is evident that $J_{sc}$ and $E_{OC}$ peak at intermediate composition, $v_r$ = 0.20. The composite film with a $v_r = 0.20$ exhibits a maximum $J_{sc} = 68.75$ nA/cm$^2$ and an $E_{OC} = 189$ V/cm under 50 N load at 5 Hz. It is worth noting that $J_{sc}$ and $E_{OC}$ generated from the $v_r = 0.20$ 2-2 PMN-0.3PT/PDMS composites are much higher than the $v_r = 0.50$ 0-3 PMN-0.3PT/PDMS composite [14], 0-3 1 vol.% CNT/30 vol.% PMN-0.3PT/PVDF composite [51], and comparable to 2-2 PZN-PZT/PDMS piezocomposites [31]. However, the decrease in the $J_{sc}$ beyond $v_r = 0.20$ is not apparent. In addition, it can be observed from Fig. 5 (c) that the $J_{sc}$ and $E_{OC}$ generated from the $v_r = 0.20$ increases with frequency.

To rationalize the above observation, FEM simulation was performed for the estimation of output voltage and current upon application of the cyclic compressive load on PGs made from 2-2 $v_r$-PMN-0.3PT/PDMS ($v_r = 0.10$, 0.20, and 0.43) parallel connectivity composites. The simulated open-circuit electric field exhibits its maximum value at $v_r = 0.20$ as shown in Fig. 6 (a), corroborating the experimental results. The FEM simulation results for the $v_r = 0.20$ composite confirm the load amplitude dependence behavior shown in Fig. 6 (b) and (c). The simulated $J_{sc}$ for $v_r = 0.20$ PGs exhibit an overestimation compared to the experimental results, as shown in Fig. 7 (a). This deviation occurs as a result of the experimental $d_{33}$ being less than 14%, which is caused by the presence of ionic defects vis-à-vis polarization leakage [38]. In addition, it can be observed from Fig. 6 (d) that the simulated $J_{sc}$ exhibits a linear increase with frequency. Increasing the frequency increases the rate of charge transfer and, consequently, the $J_{sc}$ [48]. Conversely, the simulated $E_{OC}$ for $v_r = 0.20$ PGs demonstrates an underestimation of the experimental results, as illustrated in Fig. 7 (a)..

Fig. 7 (b) depicts the influence of cyclic compressive load and frequency on the experimental and simulated power density of the piezoelectric generator (PG) based on $v_r = 0.2$, 2-2 PMN-PT/PDMS composite with parallel connectivity. It demonstrates that both the experimental and simulated power density increased with increasing compressive load and frequency [52–54]. The experimentally measured output power density for $v_r = 0.20$ composite is ~13 μW/cm$^3$ under dynamic compressive load of 50 N at 5 Hz frequency. This value is higher than the simulation result, ~10 μW/cm$^3$. The observed discrepancy in power density is mainly due to the lower open-circuit electric field predicted by the FEM simulation. The power density generated from the $v_r = 0.20$ 2-2 PMN-0.3PT/PDMS composites are higher than 0-3 (PMN-0.3PT/PDMS [14], BaTiO$_3$/PDMS [50], BaTiO$_3$/PVDF [55]) piezocomposites and comparable to 2-2 (PZN-PZT/PDMS [31]) piezocomposites. Low-powered electronic devices such as wireless sensors and portable devices can be operated with the aforementioned power [56].

## 4. Conclusion

PMN-0.3PT pellet synthesized successfully by solid-state reaction method, was used to fabricate 2-2 connectivity, $v_r$-PMN-0.3PT/PDMS ($v_r$ = 0.10, 0.20, and 0.43) piezoelectric composite. It was observed that an increase in volume fraction leads to a significant increase in the effective relative permittivity ($\kappa_{33}^c$) and elastic modulus ($E_{33}^c$). However, the effective piezoelectric charge coefficient ($d_{33}^c$) reaches its maximum value at a much lower volume fraction. The 2-2 piezoelectric composite with a $v_r = 0.20$ exhibits the highest power density among all the composites. The maximum current density, 68.75 nA/cm$^2$ and electric field, 189 V/cm, corresponds to a power density of ~13 μW/cm$^3$ was obtained from the $v_r = 0.20$ PG when subjected to a compressive load of 50 N at 5 Hz.

**Acknowledgments**

This work was funded by CSIR-HRDG, Govt. of India through project number, 22(0837)/20/EMR-II. AK thanks the Ministry of Education, Government of India, and IIT Bhubaneswar for fellowship.

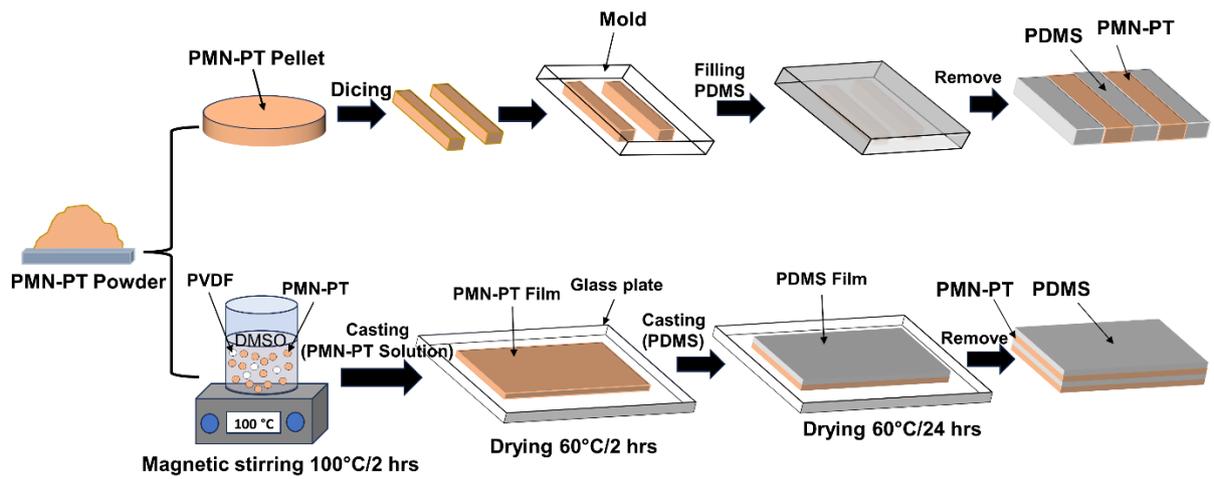

Fig. 1. The process flowchart for the fabrication of the 2-2 PMN-0.3PT/PDMS composites with both parallel and series connectivity.

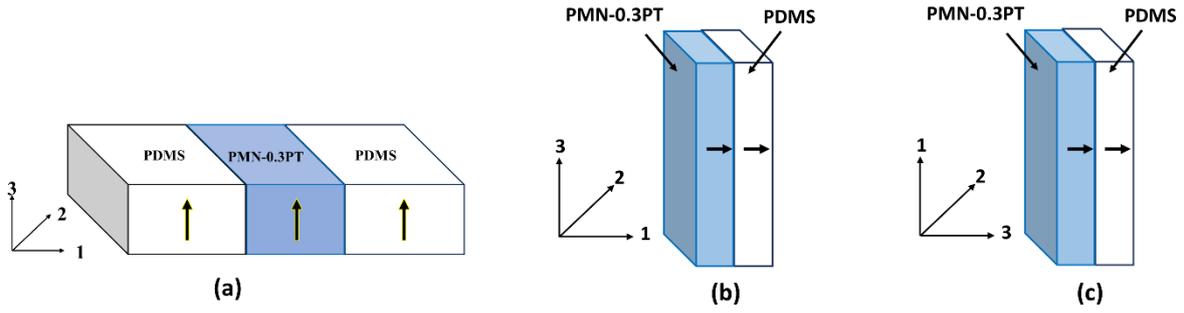

Fig. 2. (a) Multi-layered (2–2) piezoelectric composite with parallel connectivity poling along 3-direction. Multi-layered (2–2) piezoelectric composite with series connectivity poling along (b) 1-direction. (c) 3-direction.

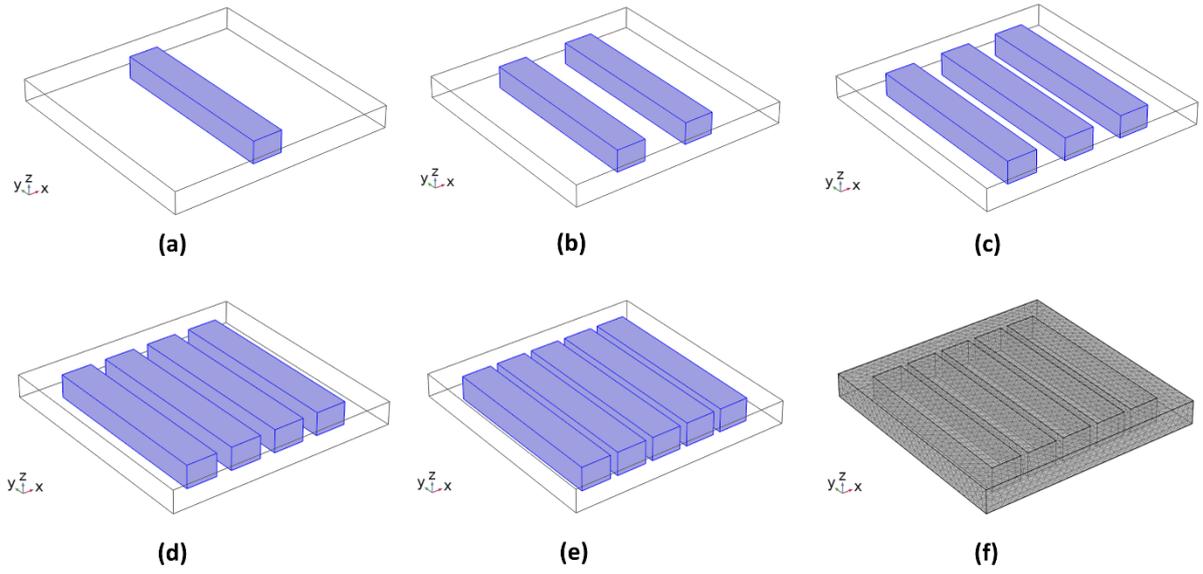

Fig. 3. 2-2 connectivity PMN-0.3PT/PDMS composite with reinforcement volume fraction of (a) 0.10 (b) 0.20 (c) 0.32 (d) 0.43 and (e) 0.53 (f) Meshed 2-2 connectivity PMN-0.3PT/PDMS composite with $v_r = 0.53$.

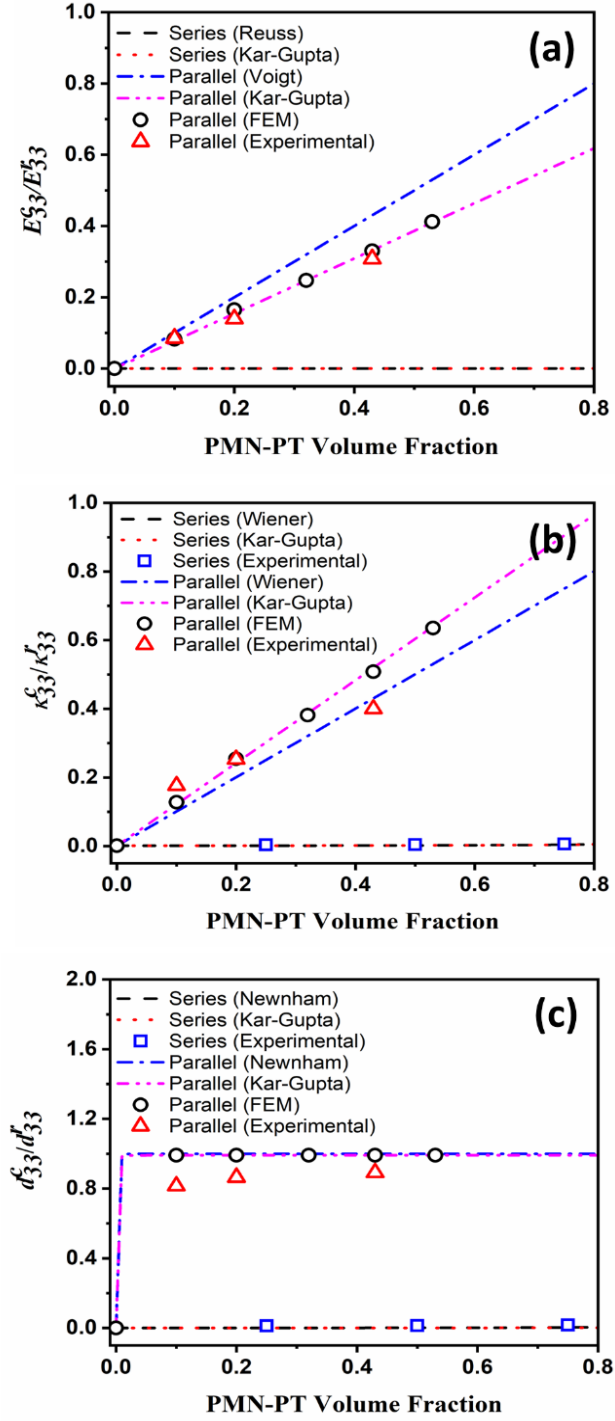

Fig. 4. (a) Comparison of normalized effective elastic modulus ($E_{33}^c/E_{33}^r$) for 2-2 PMN-0.3PT/PDMS Composite. (b) Comparison of normalized relative permittivity ($\kappa_{33}^c/\kappa_{33}^r$) for 2-2 PMN-0.3PT/PDMS Composite. (c) Comparison of normalized piezoelectric charge coefficient ($d_{33}^c/d_{33}^r$) for 2-2 PMN-0.3PT/PDMS Composite.

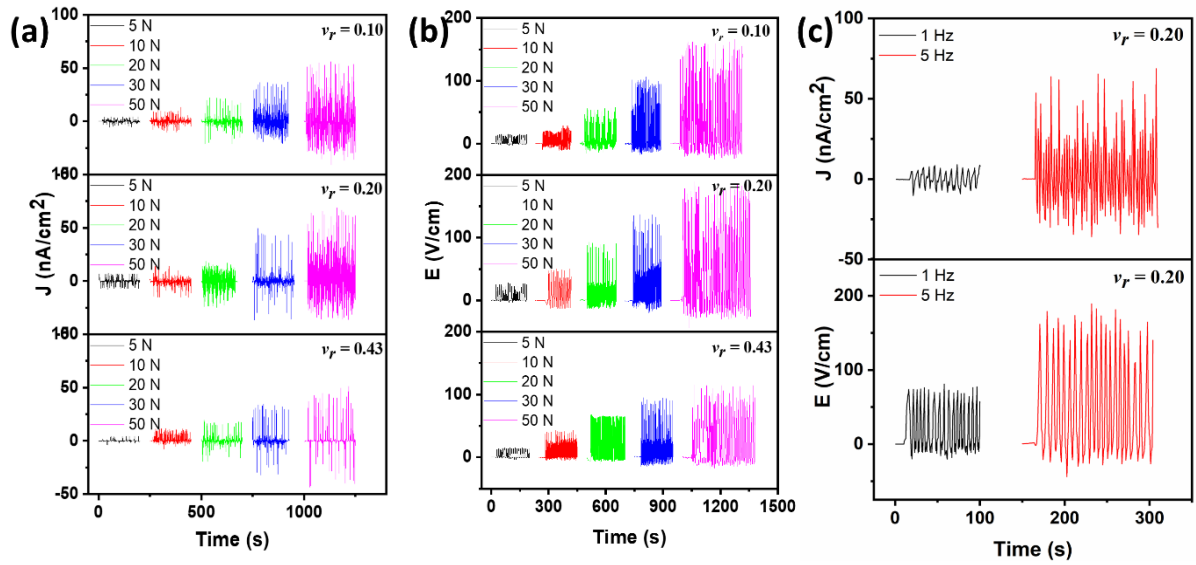

Fig. 5. (a) Variation in the short-circuit current density as a function of applied dynamic load at 5 Hz frequency. (b) Variation in the open-circuit electric field as a function of applied dynamic load at 5 Hz frequency. (c) Short-circuit current density and open-circuit electric field generated from the $v_r = 0.20$ at different frequencies at 50N.

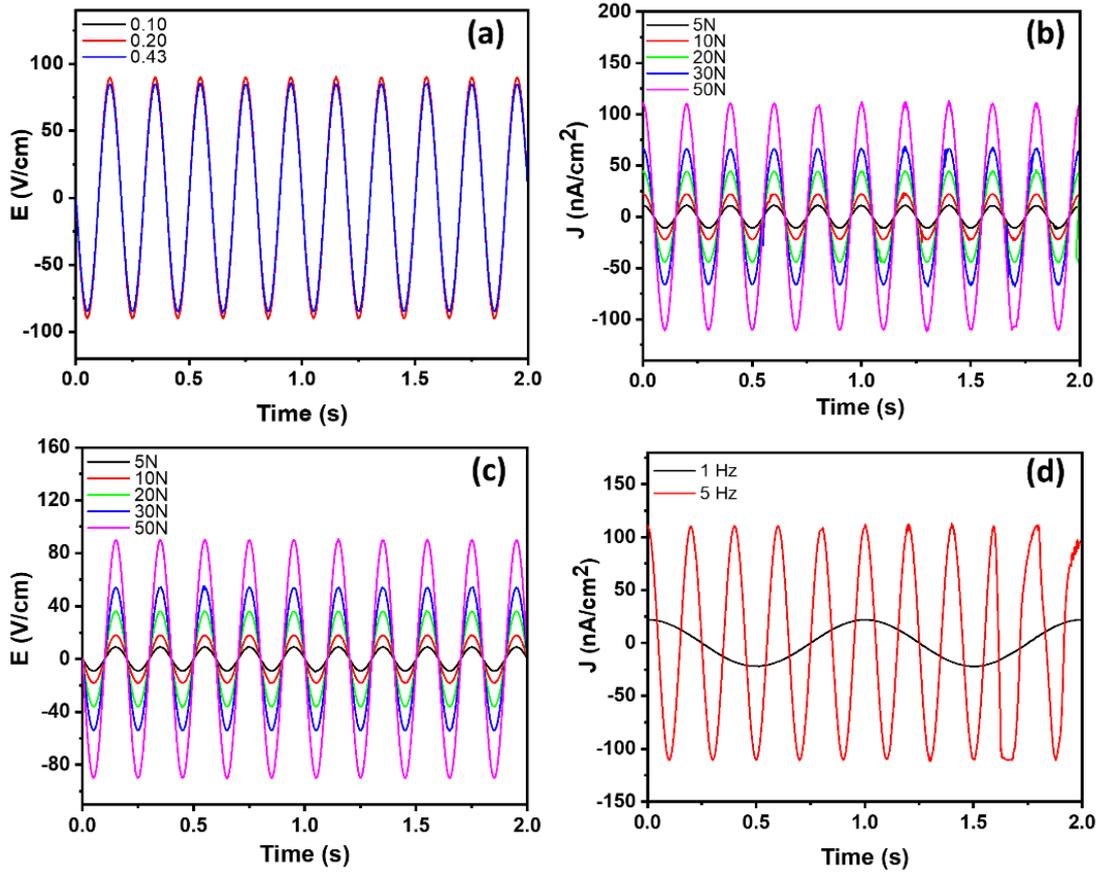

Fig. 6. Simulated results (a) Electric field response from the $v_r$ = 0.10, 0.20, and 0.43 composite at 50 N, 5 Hz. (b) Current density response from the $v_r$ = 0.20 composite as a function of load at 5 Hz frequency. (c) Electric field response from the $v_r$ = 0.20 composite as a function of load at 5 Hz frequency. (d) Current density response from the $v_r$ = 0.20 composite as a function of frequency at 50 N load.

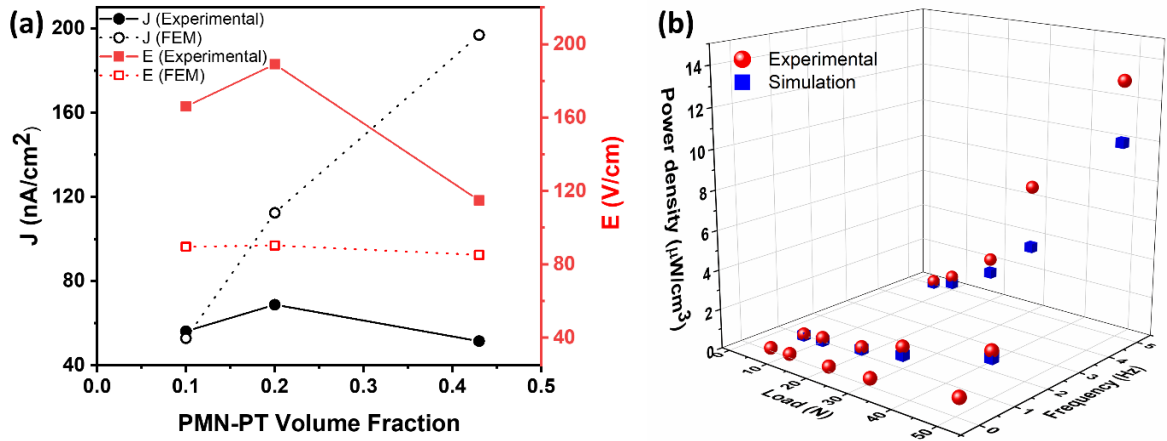

Fig. 7. (a) Experimentally measured and FEM simulation results of the output current density and electric field in response to an applied dynamic load of 50 N at 5 Hz for a PG with parallel connectivity. (b) 3D scatter plot depicts the effect of load and frequency on the experimental and simulated output power density of the $v_r$ = 0.20 piezo-generator (PGs).